\documentclass[twocolumn,appendixfloats,tighten]{aastex6}
\usepackage{newtxtext,newtxmath}
\usepackage[T1]{fontenc}
\usepackage{ae,aecompl}

\usepackage{amsmath}	
\usepackage{amssymb}	

\usepackage{ctable}
\usepackage{url}
\usepackage{xspace}
\usepackage[normalem]{ulem}
\usepackage{hyperref}
\usepackage[all]{hypcap}
\usepackage{graphicx}

\def\msun{{\rm\,M_\odot}}

\def\msun{{\rm\,M_\odot}}

\newcommand{\kms}{\, {\rm km\, s}^{-1}}

\newcommand{\be}{\begin{equation}}
\newcommand{\ee}{\end{equation}}

\newcommand{\au}{\rm au}
\def\h2{${\rm\,H_2}$}

\usepackage{color}

\begin{document}

\title{Formation and merging of Mass Gap Black Holes in Gravitational Wave Merger Events from Wide Hierarchical Quadruple Systems}
\author{Mohammadtaher Safarzadeh\altaffilmark{1}, Adrian S. Hamers\altaffilmark{2}, Abraham Loeb\altaffilmark{3}, and Edo Berger\altaffilmark{3}}

\altaffiltext{1}{Department of Astronomy and Astrophysics, University of California, Santa Cruz, CA 95064, USA,
\href{mailto:msafarzadeh@cfa.harvard.edu}{msafarza@ucsc.edu}}

\altaffiltext{2}{Max-Planck-Institut f\"{u}r Astrophysik, Karl-Schwarzschild-Str. 1, 85741 Garching, Germany}
\altaffiltext{3}{Center for Astrophysics | Harvard \& Smithsonian, Cambridge, MA 02138, USA}

\begin{abstract}
We investigate secular evolution in hierarchical quadruple systems as a formation channel of mass-gap black holes (with masses of about $3-5$ $M_{\odot}$) in systems that will eventually lead to binary black hole mergers detectable by ground-based gravitational wave detectors (LIGO/Virgo). We show that in a 3+1 hierarchical system, two episodes of induced mergers would first cause two neutron stars to merge and form a mass-gap black hole, which will subsequently merge with another (more massive) black hole through a second induced merger. We demonstrate that such systems are stable to flybys, and their formation would predict a high mass ratio and eccentric merger of a mass-gap black hole with a more massive black hole companion. Such a formation channel may explain observed gravitational wave events such as the recently-discovered LIGO/Virgo events S190814bv and S190924h.
\end{abstract}

\section{Introduction}

There have been claims in the literature for the presence of a mass-gap in the compact object mass distribution at $2-5\msun$ \citep{Ozel:2010hd,Farr:2011ct}, although more recent detailed models and observations argue for the absence of such a gap \citep{Kreidberg:2012fd,Thompson637}. This mass-gap is argued to be related to the explosion mechanism in core-collapse supernovae (SNe) prior to the formation of the black holes (BHs) in certain mass ranges \citep{Belczynski:2012dc}, and different SNe models have been proposed to justify the possible presence of a mass-gap, based on mass fall-back arguments \citep{Fryer:2011jk}. 

Although a mass-gap BH would form in the merger of two neutron stars (NSs), as has been argued in the case of the binary neutron star merger GW170817 (e.g., \citealt{Nicholl:2017}), and their assembly in dense stellar clusters \citep{2019PhRvD.100d3027R}, it would be difficult to detect such objects if they are free floating in their host galaxies \citep{Wyrzykowski:2019up}. Instead, such mass-gap BHs could be detected through gravitational wave emission if they merge with another compact object. Setting aside a random capture of a free floating mass-gap BH with a field BH and their subsequent merger due to GW emission, the only other possible way to detect them would be if the mass-gap BH is part of a binary system that would merge in less than a Hubble time. 

While evolution in a triple system might appear to be a promising channel (e.g., \citealt{2017ApJ...836...39S,2017ApJ...841...77A}), it requires fine tuning.  For example, consider a binary system of two $10\msun$ zero age main sequence (ZAMS) stars that are separated by about 1000 $R_\odot$. These two ZAMS stars would evolve through a common envelope phase and form a binary neutron star (BNS) system that would become about 1 AU apart, which is close enough to merge in less than a Hubble time.  Even if the binary separation after the common envelope phase is too large to merge within a Hubble time, a merger can be accelerated if the BNS is affected by the Lidov-Kozai (LK; \citealt{Lidov:1962du,Kozai:1962bo}; see \citealt{2016ARA&A..54..441N} for a review) effect of a third companion.  In order for the resulting mass-gap BH to merge with another compact object, we need a third massive ZAMS star companion.  However, the separation between the mass-gap BH and the third compact object companion should be less than about 6 solar radii if we assume the third companion is a black hole with mass $\gtrsim 10\msun$ and the merger is required to occur in less than a Hubble time. This requirement is almost impossible to arrange since the progenitor of a $\sim 10\msun$ BH would first expand and engulf the two ZAMS progenitors of the NSs, resulting in a common envelope phase that brings in the two ZAMS stars and the stripped stellar core of the third companion in close separation. The pre-SN compact object explodes to make a BH, and subsequently the common envelope phase of the two ZAMS will now engulf this BH. Such a system would potentially look like a Thorne-$\mathrm{\dot{Z}}$ytkow object \citep{1977ApJ...212..832T} and would act as a fast merging channel for the triple system.  
 
This situation can be avoided in a quadruple configuration. The frequency of quadruple systems depends on the primary star mass. From observations of early-type binaries, \citet{2017ApJS..230...15M} showed that for Solar-type stars, triples are $\sim 10$ times more common than quadruples; however, for systems with $\gtrsim 30\,\msun$ primary stars, triples and quadruples are about equally common, and much more common than binaries or singles. Quadruples, which are known to occur in either the 2+2 or 3+1 configuration, exhibit more complex dynamical behavior compared to triples, and generally a larger parameter space leads to strong interactions induced by secular evolution \citep{2013MNRAS.435..943P,Hamers:2014eu,2017MNRAS.466.4107H,Hamers:2017eo,2018MNRAS.474.3547G,2018MNRAS.478..620H,Hamers:2018kt,Liu:2018de,Fragione:2019ht}.

Here, we show how a hierarchical quadruple can lead to formation and merger of a mass-gap BH in a binary black hole merger event that LIGO/Virgo can detect. In \S 2 we describe the characteristics of a wide hierarchical quadruple system that would lead to a merger of a mass-gap BH with a tens of solar mass BH companion in about a Hubble time, and we provide a numerical example in \S3. In \S4, we argue that such wide systems are stable and are not disrupted due to flybys in the galactic disks. In \S 5 we discuss how to estimate the merger rate of such objects. In \S 6 we provide a summary and conclusion of our work, that detection of a mass-gap BH with LIGO/Virgo does not necessarily rule out a specific supernovae model as they could have been assembled dynamically in a hierarchical system.

\section{mass gap black hole formation in a quadruple configuration}

Our proposed scenario relies on the LK mechanism. The presence of a third companion for a binary system leads to LK oscillations between a minimum and maximum eccentricity, and in the test particle limit this occurs on a timescale \citep[e.g.,][]{Blaes:2002dj}
\begin{multline}
\label{eq:kozai_timescale}
t_{\textrm{LK}} \sim 1.8 \times 10^3 \, \textrm{yr} \left( \frac{m_1 +
m_2}{100 \, M_{\odot}} \right)^{-1/2} \left( \frac{a_1}{10^{-2}
\, \textrm{pc}} \right)^{3/2} \\
\times \left( \frac{m_1 + m_2}{2 m_3} \right) \left( \frac{a_2 / a_1}{10}
\right)^3 \left( 1 - e_2^2 \right)^{3/2},
\end{multline}
 where $m_1$, and $m_2$ are the masses of the inner binary components, and $m_3$ is the third object's mass; $a_1$, and $a_2$ are the semi-major axis of the inner and outer binary, and $e_2$ is the eccentricity of the outer binary.  

The Hamiltonian describing a triple system can be expanded in terms of the instantaneous orbital separations, which after averaging, results in expressions that contain ratios of the semi-major axis of the inner binary to the semi-major axis of the outer binary, $(a_1/a_2)$, typically to quadrupole order, namely $(a_1/a_2)^2$.  By extending the expansion to octupole order, $(a_1/a_2)^3$, it has been shown that the inner binary reaches higher eccentricities on a timescale $t_\textrm{EKM}=t_\textrm{LK}/\epsilon_{\rm oct}$ where $\epsilon_{\rm oct}$ is the strength of the octupole-order term in the expansion of the three-body Hamiltonian \citep{Naoz:2012cm,Naoz:2013ki}, although the exact timescale is debated \citep{2015MNRAS.452.3610A}. The eccentric LK mechanism timescale is 
\begin{multline}
\label{eq:tekm}
t_{\textrm{EKM}} \sim 3 \times 10^7 \, \textrm{yr} \left( \frac{m_1 +
m_2}{100 \, M_{\odot}} \right)^{-1/2} \left( \frac{a_1}{1 \,
\textrm{pc}} \right)^{3/2} \\
\times \left( \frac{m_1 + m_2}{2 m_3} \right) \left( \frac{a_2 / a_1}{20}
\right)^4 \frac{(1 - e_2^2)^{5/2}}{e_2}.
\end{multline}

The relevant timescale between Equations (\ref{eq:kozai_timescale}) and (\ref{eq:tekm}) depends on the strength of the octupole term defined as:
\begin{equation}
 \epsilon_{\rm oct} = \left( \frac{m_1 - m_2}{m_1 + m_2} \right) \left(
\frac{a_1}{a_2} \right) \frac{e_2}{1 - e_2^{2}}.
 \label{eq:epsoct}
\end{equation}
Since in our example $m_1$ is different from $m_2$, $t_{\textrm{EKM}}$ is the relevant timescale we use in this paper \citep{Shappee:2012ea,Antognini:2014ku}.

\begin{figure*}
{\includegraphics[width=\linewidth]{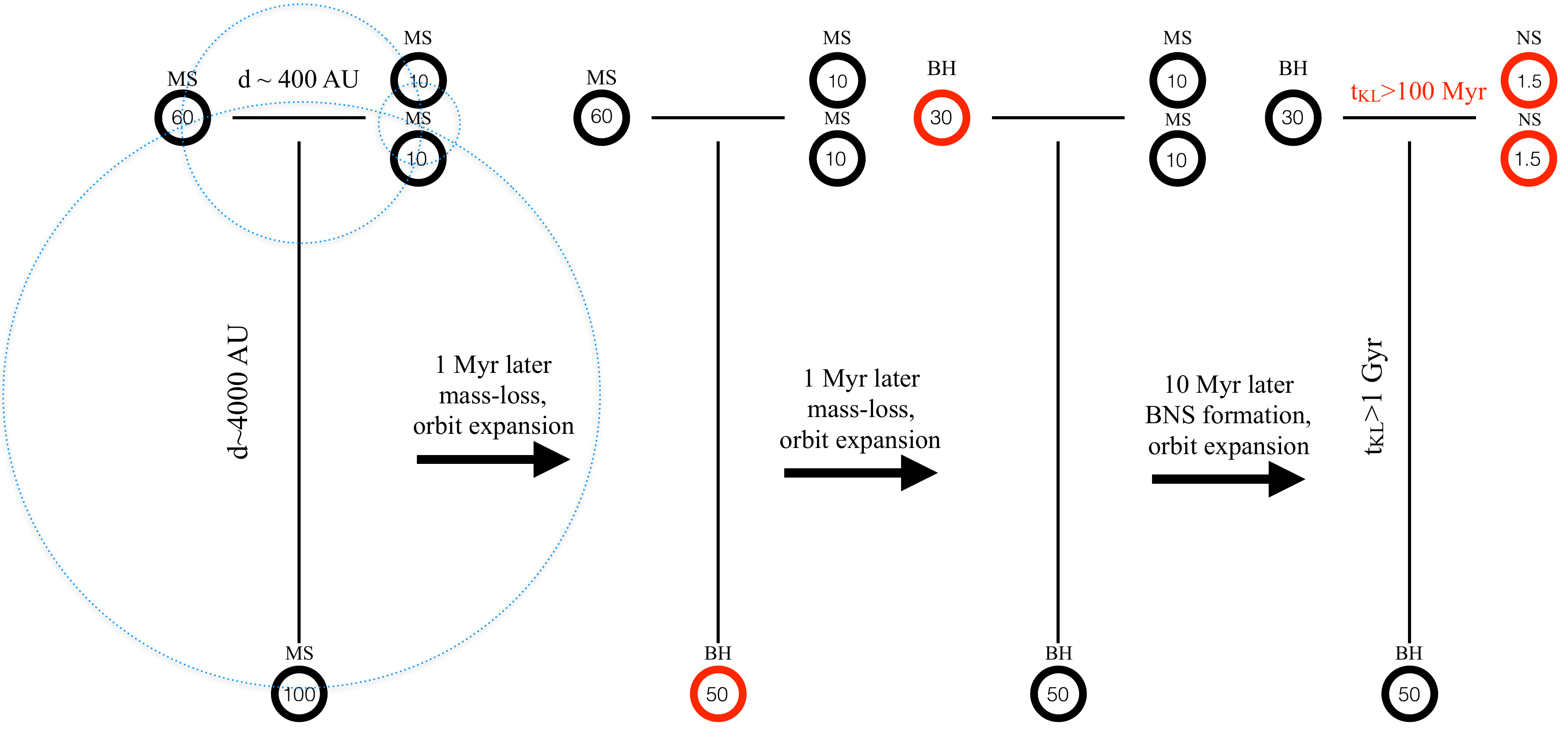}}
\caption{Wide hierarchical quadruple configuration leading to formation of a mass-gap BH and its merger with a $30\msun$ BH due to Lidov-Kozai effect. The system starts with four ZAMS stars. Starting with two $10\msun$ orbiting each other making up the innermost binary system. The innermost binary is in orbit with a $60\msun$ ZAMS which makes up the intermediate binary. The intermediate binary is in orbit with a $100\msun$ ZAMS star. The most massive stars evolve faster. The $100\msun$ make a $50\msun$ BH through a combination of mass loss and supernovae. This leads to expansion of the orbit of the outermost binary. The same happens to the $60\msun$ ZAMS which leads to the formation of a $30\msun$ BH and increase in orbital separation in the intermediate binary. The last part of the evolution is the formation of a binary NS system from the innermost binary. The two NSs merge due to LK effect of the $30\msun$ BH and form a mass-gap BH. This mass-gap BH merges with the $30\msun$ BH due to the LK effect of the $50\msun$BH.}
\label{f:fig1}
\end{figure*}

In Figure~\ref{f:fig1} we schematically show the configuration of a quadruple system that would lead to the formation of a mass-gap BH detectable in a BBH merger by LIGO/Virgo. The innermost binary system consists of two ZAMS stars of about $10\msun$. We assume the semi-major axis of this binary is $\lesssim 100$ AU, but larger than about 10 AU to avoid a formation of a BNS system through a common envelope phase. These two stars will eventually produce two NSs. The two ZAMS stars orbit a $\sim 60\msun$ ZAMS star with a semi-major axis of about 400 AU. All three stars are assumed to orbit around a $\sim 100\msun$ ZAMS star with a semi-major axis of about 4000 AU.

The evolution of such a system in our scenario is as follows. First, the most massive ZAMS star ($\sim 100\msun$) evolves and loses a fraction of its mass due to winds \citep[see ][ for a comprehensive review]{Smith:2014go}, and then explodes to form a $\sim 50\msun$ BH. We note that these numbers are not exact but are within the bounds of numerical simulations results. Due to mass loss the orbit will expand, partly adiabatically through wind mass loss, and non-adiabatically in the SN event. Second, the next most massive ZAMS star ($\sim 60\msun$) will evolve through a similar process to produce a $\sim 30\msun$ BH, and the orbit of the intermediate binary will similarly expand in response to wind and SN mass loss. Finally, the two ZAMS stars in the innermost binary will evolve on a timescale of tens of Myr to form two NSs, widening the orbit of the intermediate binary. This also impacts the outer binary, but we assume the outer binary orbit only mildly widens in this process.

In our hierarchical picture, the eccentric LK timescale of the inner binary is shorter than that of the outer binary. What happens next is that first the two NSs merge on a 100 Myr timescale which leads to the formation of a mass-gap BH. The newly formed mass-gap BH merges with the $30\msun$ BH on a timescale of about 1 Gyr due to the secular evolution. This final merger event is what we anticipate to be detectable with LIGO/Virgo as a mass-gap BH merging with a tens of solar mass BH likely with high eccentricity.

The exact evolution of a hierarchical quadruple system depends on the ratio of the LK timescales of the intermediate binary to outer binary \citep{Hamers:2014eu}. Moreover, the impact of all the non-secular effects, including the stellar and tidal evolution on a hierarchical quadruples are largely unknown \citep{Hamers:2015ku}, and therefore the next step in our study is a systematic exploration of the parameter space of binary separation and initial ZAMS masses.

\section{Numerical example}
\label{sect:num}

Here we provide a numerical example in which a double merger occurs, first between the two NSs in the innermost system producing a mass-gap BH, and then between the mass-gap BH and a more massive BH in the quadruple system.  For simplicity, we only consider the secular dynamical evolution of the system after the formation of all compact objects, i.e., we assume an initial configuration consisting of two NSs in the innermost system, and two BHs in the two outer orbits. We ignore all details of the evolution (including dynamical and stellar evolution) before the formation of the compact objects. These effects are likely important, yet we neglect them here since our objective is to illustrate that a `double merger' is in principle possible from a dynamical point of view, assuming that the system could have formed with the initial compact object configuration to begin with.

We model the secular dynamical evolution of the system using the code \textsc{SecularMultiple} \citep{Hamers:2015ku,2018MNRAS.476.4139H}, available on GitHub\footnote{\href{https://github.com/hamers/secularmultiple}{https://github.com/hamers/secularmultiple}}. This code is based on an expansion of the Hamiltonian of the system in terms of ratios of separations of adjacent orbits, assuming that the system is hierarchical. The expanded Hamiltonian is subsequently averaged over all orbits, and the resulting simplified equations of motion (compared to the $N$-body system) are solved numerically. Also included in the code are post-Newtonian (PN) terms, i.e., the 1PN and 2.5PN terms (applied to all orbital pairs), which give rise to orbital precession and orbital decay due to GW emission, respectively. Furthermore, we check in the code for the condition when an orbit becomes decoupled from its secular evolution due to GW emission, i.e., when the timescale for the orbital angular momentum to change by order itself due to secular evolution is ten times longer than the timescale for GW emission to shrink the orbit by order itself (see \citealt{2018ApJ...865....2H}, section 5.1.2). When this condition is satisfied, we stop the integration, since otherwise the integration significantly slows down\footnote{The slowdown is a result of the diverging rate of precession due to the 1PN terms as the orbit shrinks. However, this precession does not affect the evolution since the binary is already decoupled from the outer orbits when we stop the secular integration.}, and after this point in time GW emission completely dominates the evolution. 

For our numerical example, we choose the following parameters: the masses of the two NSs are $m_0 = m_1 = 1.5\,\msun$, which are orbited by a BH of mass $m_2=30\,\msun$; the latter is orbited by another BH of mass $m_3=50\,\msun$. The three orbits have initial semi-major axes $a_1=20\,\au$, $a_2=2000\,\au$ and $a_3=9000\,\au$, and eccentricities $e_1=0.01$, $e_2=0.1$, and $e_3=0.1$. The inclinations are $i_1=0.01^\circ$, $i_2=45.0^\circ$, and $i_3=146^\circ$; the arguments of periapsis are $\omega_1\simeq197.6^\circ$, $\omega_2\simeq257.5^\circ$, and $\omega_3\simeq217.0^\circ$; the longitudes of the ascending node are $\Omega_1=\Omega_2=\Omega_3=0.01^\circ$. Consequently, the initial mutual inclinations are $i_{12}\simeq 45^\circ$, and $i_{23}\simeq 101^\circ$. This type of compact object quadruple system could conceivably form as a result of stellar evolution and resulting orbital changes, starting from our prototype ZAMS quadruple system described in \S2, although, as discussed above, here we  ignore all detailed pre-compact object evolution.

\begin{figure*}
{\includegraphics[width=0.49\linewidth, trim = 10mm 0mm 5mm 0mm]{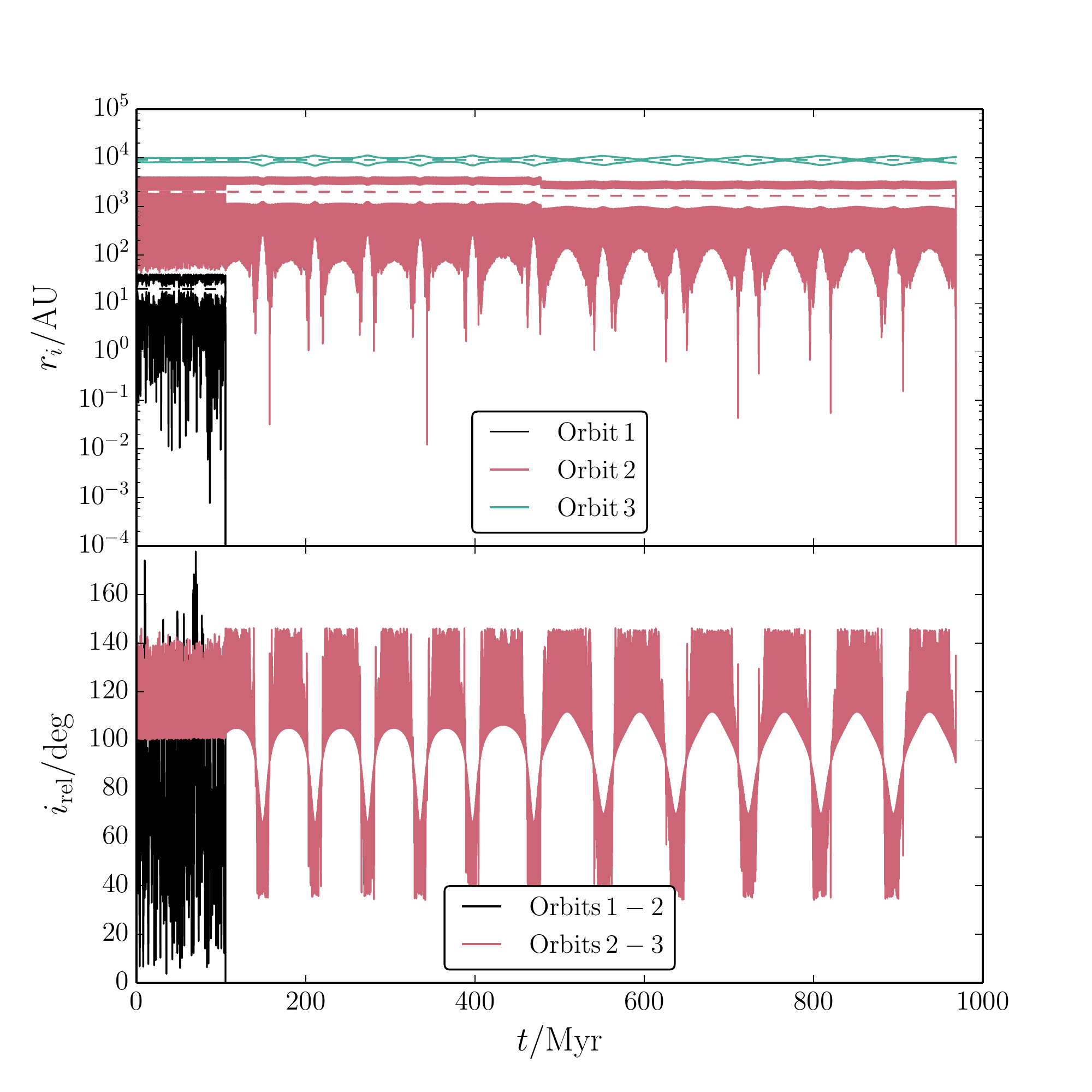}}
{\includegraphics[width=0.49\linewidth, trim = 10mm 0mm 5mm 0mm]{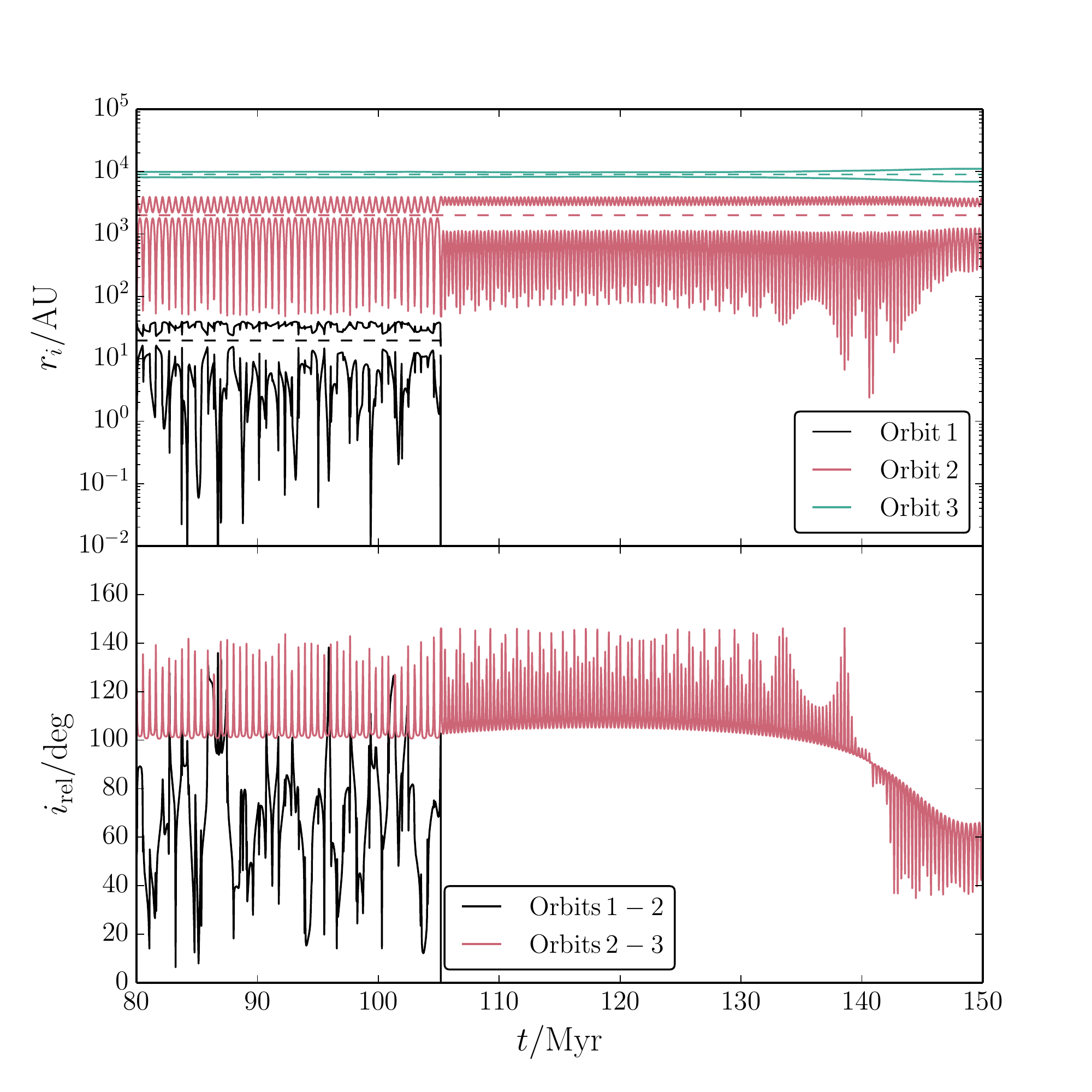}}
\caption{Numerical example of a `double LK merger'. Semimajor axes (dashed lines) and periapsis distances and apoapsis distances (solid lines) of the three orbits are shown as a function of time; the bottom panels show the mutual inclinations as a function of time. The right-hand set of panels are zoomed-in versions of the left-hand set of panels. }
\label{f:fig2}
\end{figure*}

\begin{table*}
\centering
\begin{tabular}{cccccccc}
\toprule
$m_2/\msun$ & $m_3/\msun$ & $a_2/\au$ & $a_3/\au$ & $t_\textrm{merge,first}/\textrm{Myr}$ & $t_\mathrm{merge,second}/\textrm{Myr}$ & $e_\textrm{LIGO,first}$ & $e_\textrm{LIGO,second}$ \\
\midrule
29.2 & 35.6 & 1703.3 & 15058.6 & 75.0 & 2936.0 & 0.124 & 0.0039 \\
23.6 & 32.0 & 1997.8 & 15445.1 & 215.0 & 9828.0 & 0.2376 & 0.0034 \\
44.7 & 46.8 & 1851.2 & 13691.4 & 55.0 & 4151.0 & 0.1086 & 0.0026 \\
46.0 & 43.5 & 1967.4 & 14970.2 & 274.0 & 2559.0 & 0.2127 & 0.0052 \\
26.6 & 47.6 & 1924.4 & 14264.6 & 66.0 & 3063.0 & 0.3081 & 0.0042 \\
27.4 & 46.4 & 1915.7 & 13104.3 & 131.0 & 385.0 & 0.305 & 0.012 \\
32.5 & 44.9 & 1917.6 & 13499.6 & 37.0 & 2099.0 & 0.2778 & 0.0048 \\
36.2 & 46.3 & 1830.9 & 14327.8 & 45.0 & 1133.0 & 0.259 & 0.0075 \\
47.7 & 34.4 & 1974.3 & 15993.5 & 172.0 & 1622.0 & 0.3816 & 0.0005 \\
37.3 & 44.4 & 1908.3 & 14928.8 & 147.0 & 1819.0 & 0.0958 & 0.0063 \\
24.7 & 32.3 & 1979.4 & 13386.8 & 116.0 & 2277.0 & 0.1541 & 0.0112 \\
31.2 & 37.0 & 1880.4 & 13027.3 & 41.0 & 270.0 & 0.257 & 0.0091 \\
39.6 & 43.5 & 1953.3 & 14514.4 & 175.0 & 1140.0 & 0.1278 & 0.0069 \\
47.9 & 29.9 & 1952.3 & 13539.9 & 208.0 & 1249.0 & 0.4193 & 0.0034 \\
45.4 & 49.1 & 1981.7 & 15704.1 & 80.0 & 3510.0 & 0.1053 & 0.0042 \\
\bottomrule
\end{tabular}
\caption{Properties of double-merger systems found in our restricted set of simulations (see Section~\ref{sect:num}). All systems (out of 1000 sampled systems) in which a double merger occurred are listed; the first four columns show the corresponding sampled parameters $m_2$, $m_3$, $a_2$, and $a_3$. The next two columns show the time of first merger in the innermost orbit of the quadruple system ($t_\textrm{merge,first}$) and of the second merger in the inner orbit of the subsequent triple system ($t_\mathrm{merge,second}$; measured with respect to the same zero point as $t_\textrm{merge,first}$). The last two columns show the associated eccentricities when entering the LIGO band at $f_\textrm{GW}=10\,\textrm{Hz}$, $e_\textrm{LIGO,first}$ and $e_\textrm{LIGO,second}$ for the first and second merger, respectively. }
\label{table:set}
\end{table*}

We show the secular evolution of the compact object quadruple system in Figure~\ref{f:fig2}; the top panels show the semimajor axes (dashed lines) and periapsis distances and apoapsis distances (solid lines) of the three orbits as a function of time (measured from the time of compact object formation); the bottom panels show the mutual inclinations as a function of time.  During the first $\sim 100\,\textrm{Myr}$, the innermost system evolves chaotically in its eccentricity despite the initial modest mutual inclination with respect to the second orbit, which is characteristic for quadruple systems (e.g., \citealt{Hamers:2014eu,Hamers:2017eo,2018MNRAS.474.3547G}). Very high eccentricities are reached when the inclination relative to the second orbit, $i_{12}$, switches values between prograde and retrograde orientations, and vice versa. At $\lesssim 105\,\textrm{Myr}$, the NS-NS binary reaches sufficiently high eccentricity that it becomes decoupled from the secular perturbations from its companions, and it rapidly merges due to GW emission in $\sim 6\,\textrm{yr}$. The orbit is sufficiently eccentric at the moment of decoupling ($e_1 \simeq 1-1.1\times10^{-6}$ with a semimajor axis of $a_1\simeq 8.16\,\au$), that a significant eccentricity still remains when the orbit reaches the LIGO frequency band at $10\,\textrm{Hz}$, i.e., $e_1\gtrsim 0.19$ when $f_\textrm{GW}=10\,\textrm{Hz}$\footnote{Here, we calculated $f_\textrm{GW}$ using Equation 37 of \citet{2003ApJ...598..419W}.}. 

Before the innermost binary merged, it induces precession on the intermediate orbit, which in turn partially suppresses LK oscillations induced by the fourth body. This effect \citep{Hamers:2014eu} also occurs, for example in circumbinary planet systems in stellar triples \citep{2016MNRAS.455.3180H}.  After the merger of the innermost binary, however, its `shielding' effect disappears, and the intermediate orbit becomes more susceptible to high-eccentricity LK oscillations induced by the fourth body. Specifically, octupole-order effects become important and they are manifested in Figure~\ref{f:fig2} as longer-timescale modulations of the shorter-timescale quadrupole-order oscillations. After $\lesssim 1\,\textrm{Gyr}$, the orbit of the mass-gap BH and the $30\,\msun$ BH becomes eccentric enough for it to be decoupled from the secular excitations induced by the $50\,\msun$ BH, and it merges due to GW emission within $\sim 200\,\textrm{yr}$. The orbit is again highly eccentric, with $e_2\simeq 1-2.5\times10^{-7}$, and $a_2 \simeq 329\,\au$. Note that some dissipation of orbital energy occurred before decoupling. There is a small remaining eccentricity in the LIGO/Virgo band in this case, i.e., $e_2\simeq 0.0048$ at $f_\textrm{GW}=10\,\textrm{Hz}$.

This numerical example shows that it is in principle possible for a `double merger' to occur in 3+1 quadruple systems. We note that the system is inherently chaotic, especially with respect to the innermost orbit before the first merger. Slightly changing the initial conditions will affect the outcome, and hence fine-tuning is required for a double merger to occur. A larger number of detailed simulations is therefore needed to obtain a reliable estimate of the probability for a double merger given initial distributions of the orbital parameters of the system. We emphasize that our example system is a proof of concept of the secular dynamical evolution only, and we completely ignored the pre-compact object evolution. Another caveat is that we employed the secular approximation, which can break down in compact-object systems (e.g., \citealt{2012ApJ...757...27A,2014ApJ...781...45A}). 

Although a comprehensive study taking into account the above caveats is beyond the scope of this paper, we briefly discuss results from a set of restricted simulations to illustrate that, although a double merger requires a fine-tuned system to occur, the degree of fine-tuning needed is not implausible.

In this restricted set of simulations, we assume the same parameters as in our numerical example above (i.e., corresponding to Figure \ref{f:fig2}), but vary the two masses $m_2$ and $m_3$ and the two semimajor axes $a_2$ and $a_3$ through Monte Carlo sampling. Specifically, we sample both $m_2$ and $m_3$ from a flat distribution in mass between 20 and 50 $\msun$, whereas we sample the semimajor axes $a_2$ and $a_3$ from flat distributions in $\log_{10}(a_2)$ and $\log_{10}(a_3)$, with $1000\,\au<a_2<2000\,\au$, and $1.2\times 10^4\,\au<a_3<1.6\times 10^4\,\au$. We note that the mass distributions of BHs are (still) not well constrained, especially in quadruple-star systems. A flat distribution in the log of the semimajor axis is the well-known \"{O}pik distribution \citep{1924PTarO..25f...1O}; here, we ignore any deviations from this distribution in our quadruple systems of compact objects due to different masses considered (e.g., \citealt{2012Sci...337..444S}), and the impact of high multiplicity and stellar evolution (e.g., \citealt{2018MNRAS.478..620H}). We integrate the quadruple system for 1 Gyr; if a merger occurs during this time, we integrate the subsequent triple system for another up to 14 Gyr.

Out of 1000 systems sampled according to the above description, we find 15 double mergers (and hence mass-gap mergers). Some properties of these systems (time of merger, and eccentricity when entering the LIGO band) are listed in Table~\ref{table:set}. By construction, the time of first merger is $\leq 1\,\textrm{Gyr}$. Most first mergers in these systems occur before $\sim$ few 100 Myr. The second merger typically takes longer to occur (up to $\sim$ 10 Gyr), as can be expected from the longer secular evolution timescales in the triple system after the first merger. In all cases, the eccentricity when entering the LIGO band is larger for the first merger compared to the second.

The fraction of double mergers (1.5\%) in our restricted sample is small, yet measurable within a modest sample of 1000 systems. Although the restricted sample only represents a small fraction of the complete population of quadruples, this illustrates that, although double mergers (and, hence, mass-gap mergers from quadruples) require finely-tuned systems, the extent of fine-tuning is not implausibly high.

\section{Stability of a wide quadruple system}

An additional concern about a wide 3+1 system is its stability, as wide binaries tend to be disrupted by perturbers in the disks of their galaxies \citep{Binney:2008wd}. The timescale for tidal disruption of a wide binary in the diffusive regime --- defined as when the passage of the perturber is not close to the semi-major axis of the binary --- is given by:
\be
\frac{t_{\rm diff}}{\rm Gyr}=\left(\frac{v_p}{200\kms}\right) \left(\frac{100 \msun}{M_p}\right) \left(\frac{0.01 \msun pc^{-3}}{\rho_p}\right) \left(\frac{0.1 {\rm pc}}{a}\right) \left(\frac{M_b}{\msun}\right)
\ee
and in the catastrophic regime where the system is disrupted by single, closest encounter, as:
\be
t_{\rm cat}= 3  \left(\frac{0.01 \msun pc^{-3}}{\rho_p}\right) \left(\frac{0.1 {\rm pc}}{a}\right) ^{3/2}\left(\frac{M_b}{\msun}\right)^{1/2} {\rm Gyr}
\ee
with the transition taking place for perturber of mass above $m_{\rm crit}$ defined as:
\be
m_{\rm crit}=30\msun  \left(\frac{v_p}{200\kms}\right) \left(\frac{a}{0.1 {\rm pc}}\right)^{1/2} \left(\frac{M_b}{\msun}\right)^{1/2}.
\ee
Here $M_b$ is the total mass of the binary, $v_p$ is the perturber's velocity, and $a$ is the semi-major axis of the binary. Given the characteristics of our quadruple system, the outer binary mass in the final step that consists of two NSs and two BHs, is about $M_b\approx100\msun$. assuming the semi-major axis of the outer binary is about 0.1 pc after all the expansions due to mass losses, and setting $v_p=20\kms$ to approximate the velocity dispersion of the stars in the disk, we arrive at $m_{\rm crit}\approx30 \msun$. However, the perturbers of our binary systems are mostly the most abundant stars in the disk that have the mass of about $0.6\msun$. Therefore, the relevant timescale for the disruption would be the diffusive regime, which would be on the order of $\sim$30 Gyr given our parameters. 

\section{Merger Rate}

In this study we did not model an ensemble of quadruple systems and follow their evolution by direct numerical integration as is done in previous studies \citep[e.g., ][]{Fragione:2019ht}. Our goal was to sketch out a new possible path for the formation of a mass-gap BH in merging BBH systems without relying on specific supernovae mechanism to explain their existence. However, here we provide an order of magnitude estimate of the birth rate and detectability of such systems by LIGO.

The LIGO detection horizon for BNS mergers is  $\approx$200 Mpc, scaling as $\propto M_{c}^{5/6}$ where $M_c$ is the chirp mass of the system. In our model the expected chirp mass for a system of a mass-gap BH and a massive BH of about 30 $\msun$ is $M_c=7.5\msun$ which increases the detection range of LIGO out to 1 Gpc. The current detection rate of the BBH mergers is $\approx 50~\rm Gpc^{-3}\rm yr^{-1}$ and any source with merger rate above $0.01~\rm Gpc^{-3}\rm yr^{-1}$ has a considerable chance of detection within the next decade.

If we adopt an average star formation rate for the universe to be $10^{8}\msun~\rm Gpc^{-3}\rm yr^{-1}$ \citep{Madau:2014gtb}, a putative formation efficiency of one such quadruple system per $10^{10}\msun$ leads to a birth rate of $0.01\msun~\rm Gpc^{-3}\rm yr^{-1}$. Given that BBH formation efficiency is about one per $10^6-10^7\msun$ \citep{2019ApJ...883L..24S}, our quadruple channel scenario can have $10^3-10^4$ times lower formation efficiency than BBHs, yet result in one such merger detectable with LIGO within the next decade.

Given that the abundance of the quadruple systems at high primary mass range exceeds that of binary systems, for this channel to lead to an observable mass-gap BH merger detectable with LIGO, we require 0.01-0.1\% of all the quadruple systems formed to have favorable configuration and survive the non-secular evolution including the inner binary neutron star natal kicks. 

A mass-gap BH merger takes place in about 2\% of the time relative to the restricted parameter space we simulated. Therefore, once we form the final quadruple system consisted of four compact objects, for this channel to work, we require at least a 1\% chance of surviving the stellar evolution process leading to the formation of the compact objects. Such evolutions have been studied in the case of triple systems \citep{Hamers:2013cp,2016ComAC...3....6T,2016MNRAS.460.3494S}. Since we have assumed that the initial separations of the ZAMS are wide enough, such that the system avoids a common envelope phase, the most disruptive stellar evolution would be the natal kicks of the newly born neutron stars. 
We have assumed the BHs form with no natal kicks given their large masses. If there is a BH natal kick, the outer binary can be easily disrupted with only a few km/s kick. Since our knowledge of the quadruple frequency is incomplete, it suffices to illustrate that if 0.01-0.1\% of all the quadruples fall into our formation channel, they have a chance of reproducing a mass-gap merger event detectable with LIGO.
Lastly, we note that we consider it unlikely for the inner binary to merge while the stars are on their main sequence, as mass loss and the subsequent widening of the orbit will make the inner binary more susceptible  to LK oscillations \citep{Shappee:2012ea,2016MNRAS.462L..84H,2017ApJ...844L..16S}.

If a mass-gap event in LIGO is formed through our proposed model, we expect the event to have two main characteristics: (i) high eccentricity, and (ii) large mass ratio.
Assisted merging of BBHs through LK mechanism has been suggested in literature. For example, \citet{2018ApJ...856..140H} shows that the merger rate of BBHs around central massive BHs 
could be between 1-3 $\rm Gpc^{-3} year^{-1}$. It is plausible to assume that a triple system around a massive central BH would lead to a similar signal, where first the two NSs merge in the triple configuration forming a mass-gap BH, which then merges with the third body due to the induced LK of the central massive BH. 
This mechanism would have a rate reducution depending on the relative frequency of triples to binaries, and we expect this to be comparable to our proposed model, although more detailed 
calculation would need to be performed to show this explicitly. However, we do not expect such systems to form in a dense stellar system due to mass segregation. Therefore, mechanisms proposed to form compact binaries in globular clusters \citep[e.g., ][]{2018PhRvL.120s1103K} would not lead to formation of a high mass ratio event, such as the one we propose in this work.

\section{Conclusions \& Summary}

The LIGO/Virgo Scientific Collaboration announced detections of mergers with a possible mass-gap component (S190814bv and S190924h), implying a class of GW merger events where at least one of the BHs has a mass of $3-5\msun$. The actual masses of the binary components have not been published yet, but we predict in both cases that the binary has a high mass ratio, with the more massive component having a mass of $\gtrsim 10\msun$ and possibly much higher. We argue that such a system could be formed in a wide hierarchical quadruple system as illustrated above, and therefore, such a discovery does not necessarily inform us about SN explosion mechanisms. We show that a hierarchical quadruple system in a 3+1 configuration is able to form a mass-gap BH and provide the conditions for it to merge with another BH such that LIGO will be able to detect it \citep[most likely as an eccentric system, assuming the head-on collision fraction is sub-dominant; ][]{Hamers:2013cp}. 

In this study we focused on a 3+1 system as when the innermost system merges, the merged body still forms a triple with two other bodies, and secular evolution could accelerate the merger of the newly-formed inner binary. 
On the other hand, in 2+2 systems, when the two innermost binaries merge, they form a new binary which would likely be wide, and probably not merge in isolation within a Hubble time. 
Although, if it would happen, perhaps aided by some fortuitous flybys, one could in principle get a merger of two mass-gap BHs if the system originally consisted of four NSs.
\\
\acknowledgements 

We are thankful to the referee for their constructive comments. 
MTS is thankful to Will M. Farr, Enrico Ramirez-Ruiz, and Selma E. de Mink for helpful discussions. 
MTS is grateful to the Harvard-Smithsonian Center for Astrophysics for hospitality during the course of this work. 
This work was supported in part by the Black Hole Initiative at Harvard University, which is funded by JTF and GBMF grants.

\end{document}